\documentclass[final]{svjour3}
\usepackage{graphicx}
\usepackage{rotating}
\usepackage{amssymb}
\usepackage{mathptmx}
\usepackage[numbers]{natbib}
\makeatletter
\journalname{Journal of Low Temperature Physics}

\bibpunct{}{}{,}{s}{}{,}

\begin{document}

\newcommand{\hdblarrow}{H\makebox[0.9ex][l]{$\downdownarrows$}-}
\title{Generation of vortices and observation of Quantum Turbulence in an oscillating Bose-Einstein Condensate}

\author{E.A.L. Henn$^1$ \and J.A. Seman$^1$ \and G. Roati $^2$\and K.M.F. Magalh\~{a}es$^1$ \and V.S. Bagnato$^1$}

\institute{1:Instituto de F\'{\i}sica de S\~ao Carlos, Universidade de S\~ao Paulo,\\ Caixa Postal 369, 13560-970 S\~ao Carlos, SP, Brazil\\
Tel.:+55 16 3373 9823\\ Fax:+55 16 3373 9811\\
\email{ehenn@ifsc.usp.br}
\\2: LENS and Dipartimento di Fisica, Universita di Firenze, and INFM-CNR, \\Via Nello Carrara 1, 50019 Sesto Fiorentino, Italy\\}

\date{15.06.2008}

\maketitle


\begin{abstract}

We report on the experimental observation of vortex formation and production of tangled vortex distribution in an atomic BEC of $^{87}$Rb atoms submitted to an external oscillatory perturbation. The oscillatory perturbations start by exciting quadrupolar and scissors modes of the condensate. Then regular vortices are observed finally evolving to a vortex tangle configuration. The vortex tangle is a signature of the presence of a turbulent regime in the cloud. We also show that this turbulent cloud has suppression of the aspect ratio inversion typically observed in quantum degenerate bosonic gases during free expansion. 

\end{abstract}

\section{Introduction}

Superfluidity and the properties associated with it like quantized vortices and turbulence, or Quantum Turbulence (QT), have been extensively studied in superfluid Helium below the Lambda point both experimentally and theoretically since it has been discovered about 50 years ago \cite{Hall}. QT is a phenomena characterized by the appearance of quantized vortices distributed in a tangled way \cite{QT, QT2, tsubotaturb}. Until recently, turbulence in He-superfluid could only be observed by indirect methods \cite{Hemethod}, since individual vortices are too small to be observed directly. A few years ago QT in Helium \cite{hydrogen} has been directly observed by means of imaging solid particles or atoms trapped in the vortices cores. Theoretically, QT in Helium has been satisfactorily modeled by the vortex filament model \cite{vfm}, though some features such as vortex reconnections must be introduced artificially. Up to very recently, Helium was the only superfluid providing evidences of QT and revealing its properties. 

The achievement of BEC in trapped atomic gases \cite{bec} and the subsequent observation of quantized vortices in these samples \cite{vortex} opened up the possibility to study turbulence in a more controlled fashion, since one can control the main characteristics (interaction, atomic density, number of atoms, trapping configuration) of the atomic cloud. In fact, the study of QT in atomic superfluids may shine light on the turbulence characteristics that are due to the superfluidity as a macroscopic quantum state and reveal which of them are in close connection to the superfluidity itself. 

Quantized vortices in BECs were preferentially produced by inducing rotation in the superfluid atomic cloud around a single axis by means of rotating asymmetric traps \cite{assymetric potential} and/or stirring potentials \cite{madison}. That methods have little chances to produce QT because the single sense of rotation and consequent regularity with which the vortices are nucleated. In fact, it has been shown experimentally \cite{abo} and theoretically \cite{vortexsimul} that rotating a BEC in a single direction generate vortices that evolve to the well-known vortex lattices. Recently, vortex formation has been also observed by merging multiple trapped BECs \cite{Scherer} and by coherent transfer of orbital angular momentum from optical fields to the condensed sample \cite{LG}.

In this communication, we report on the production of multidirectional vortices which allowed the experimental observation of QT in a magnetically trapped BEC of $^{87}$Rb atoms evidenced by the observation of tangled vortices in the quantum sample. Following the identification of QT, a dramatic change in the hydrodynamic behavior of the atomic cloud during free expansion was revealed. The usual aspect ratio inversion of the condensate is suppressed and the turbulent cloud expands keeping its anisotropic initial spatial profile. We strongly believe that this behavior is intrinsically associated to the presence of turbulence in the superfluid. That effect resembles a kind of self-trapping due to the tangled vortices configuration. 


\section{Experimental Procedure}

\begin{figure} 
\centering
 \includegraphics[scale=0.4]{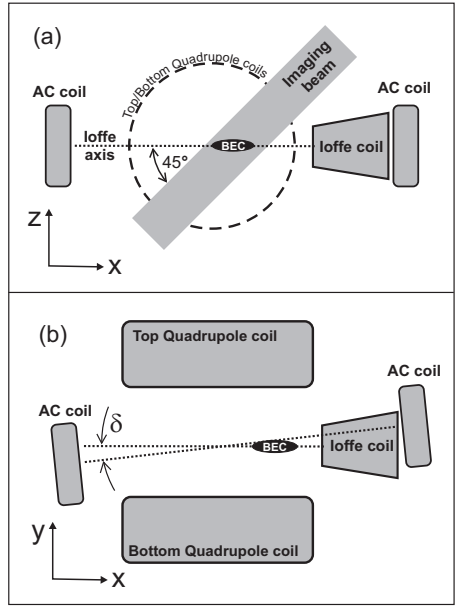}
 \caption{Schematics of (a) above view, and (b) side view of the imaging beam and the coil arrangement in our system, showing the QUIC trap coils and the excitation coils (AC coil). The drawing is not to scale and the excitation coils misalignment is made bigger for clarity.}
\label{fig:oaccoil}
\end{figure}

The description of the experimental sequence to achieve Bose-Einstein Condensation as well as the protocol to generate vortices in the condensate are described in details in Refs.\cite{BJP,vortexform, PRL}. In brief, we produce a BEC of $^{87}$Rb containing $\left(1-2\right)\times10^5$ atoms in a cigar-shaped magnetic trap with frequencies given by $\omega_r=2\pi\times210 (3)$ Hz and $\omega_x=2\pi\times23 (3)$ Hz. We measure our trapping frequencies using the standard procedure with cold atoms. We induce simple dipolar motion of the condensate by rapidly changing the bias field of the trapping potential, let the condensate evolve freely and measure its position as a function of time. That movement is periodic and its frequency is the trapping frequency. Recently, we have published a theoretical analysis \cite{caracanhas} of the expansion dynamics of our condensate, when it coexists with a large thermal component, and with those above mentioned trapping frequencies as input for the model, the experimental data can be quite well modeled. Fig.\ref{fig:oaccoil}(a) illustrates the experimental setup, showing the trapping coils and absorption image beam. 

After reaching BEC, an extra field, produced by a pair of anti-Helmholtz coils is superimposed on the magnetic trap field as shown in Fig.\ref{fig:oaccoil}(b). The axis of the extra coils is close to the trap axis (the angle between these axes is $\delta\sim 5^o$). The center of the extra field, defined by the zero-field amplitude position, is close to the QUIC trap minimum. An oscillatory current is applied to the coils, always having the same sign and always starting from zero, so we do not give an abrupt kick to the condensate in the beginning of the excitation phase ($I_{coil}=I_0\left[1-\cos(\Omega t)\right]$). The anti-Helmholtz coils produce a quadrupolar magnetic field, given by $\vec{B}=A\left(-2x\hat{x'}+y\hat{y'}+z\hat{z'}\right)$, where A is the magnetic field gradient and $\hat{x'}$ is the direction parallel to the line that join the center of the coils. 

The experimental sequence is then as follows: finished the evaporative cooling, the external field is turned on from zero to 100 ms. For this communication the frequency of oscillation has been kept constant at 200 Hz and the amplitude of the magnetic field gradient has been varied from zero up to $190$ mG/cm. After the end of the oscillation stage, atoms are left trapped for extra 20 ms before being released and observed in free expansion by a standard absorption imaging technique. 

\section{Results}
\subsection{Vortex nucleation and emergence of turbulence}

\begin{figure}
\begin{center}
\includegraphics[%
  width=0.5\linewidth,
  keepaspectratio]{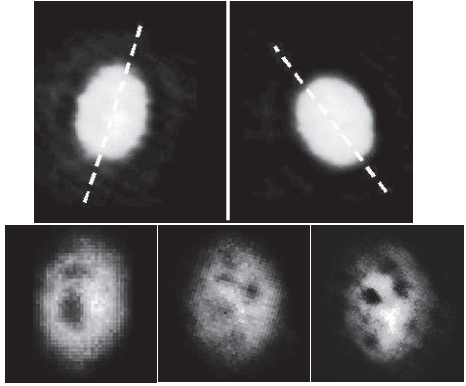}
\end{center}
\caption{(Color online) Top line: evidences of scissors modes (bending of the symmetry axis) after low-amplitude excitation by the external magnetic field. Bottom line: from left to right, increasing number of vortices observed in the BEC as the amplitude of the excitations is increased. All images taken after 15 ms of free expansion. }
\label{2}
\end{figure}

As the oscillations are introduced in the cloud, for small amplitudes of the oscillating field as well as short excitation periods we observe dipolar, quadrupolar and scissors modes of the BEC. The importance of this experimental observation is that the excitation of these modes give evidence that our extra field disturbs mechanically the cloud, changing not only its rest position but also its symmetry axis and shape by means of changing the trapping frequencies. The condensed and thermal clouds do not follow these mechanical disturbances in the same way giving rise to a relative movement between them. That relative movement is a fundamental feature for the possible mechanism of vortex formation that we briefly discuss below. 
Increasing both parameters (time of excitation and amplitude) we start to observe the formation of vortices. They seem to be formed preferably on the edges of the quantum fluid. The number of vortices observed grows as a function of the amplitude and/or duration of excitation. Examples of such regimes are shown in Fig.\ref{2}. The formation dynamics of the vortices as a function of the amplitude of the excitation field has been extensively discussed in Ref.\cite{vortexform}.

When the amplitude and/or excitation time is increased above 160 mG/cm a completely different regime takes place. We observe an increase in the number of vortices followed by a proliferation of vortex lines not only in the original plane of the individual vortices, but distributed for the whole sample, covering many directions. An example of such a regime can be seen in the images of the right column of Fig.\ref{4}a.
 
The presence of quantized vortices non regularly distributed along the whole sample in all directions is quite characteristic of turbulence in the quantum fluid and it is here taken as evidence for this regime. In fact, our way of nucleating vortices, where non-equivalent translational and torsional movements of the quantum cloud occur in different planes of symmetry of the sample, can be seen as analogous to what is proposed in \cite{tsubotaturb} for combined rotations. 

Compared to liquid Helium, the total number of vortices contained in the cloud is quite limited. The relative low density of the atomic fluid together with its finite size cloud creates a limit for the maximum obtained number of vortices that is much smaller than in liquid Helium. Consequently some turbulence features might not be possible to be observed and/or be very different compared from that obtained in liquid Helium. In this class are effects related to large volumes cases of quantum turbulence, containing many vortices, specially those related to the Komolgorov spectra, where large volumes are needed. On the other hand, the low vortex density may allow one to make detailed observations of vortex reconnections and all its related phenomena such as cascade-like processes. In order to test the Komolgorov spectra, the Bragg scattering technique used by Muniz \textit{et al.}\cite{Muniz} to study vortex lattices is probably the best candidate. In the case of observation of turbulence decay, in-situ, non-destructive images should be used \cite{insitu}.

Concerning the vortex formation during oscillating perturbations, we believe that a possible mechanism of vortex formation are the so-called Kelvin-Helmholtz instabilities \cite{KH}, a well-known phenomena in the scope of fluid interface theory and experiments. These instabilities occur at the interface of two fluids with a relative motion and give rise to the formation of vortices. This phenomenon has been theoretically investigated \cite{KH theory} in normal fluid-superfluid mixtures and directly observed in mixtures of A and B phases of superfluid liquid Helium \cite{KH superfluid}. Nevertheless, it has never been observed in atomic BEC, though we have shown strong evidence of it in a previous work \cite{vortexform}. In our case, the superfluid is the BEC component whereas the normal fluid is the thermal cloud. Both clouds have different frequencies of oscillation for quadrupolar and scissors modes, giving rise to a relative movement and consequent friction between them. Since the relative movement is periodic, we believe that at this point vortices and anti-vortices are nucleated, leading the cloud through the regular vortices (with possible presence of anti-vortices) regime until it reaches the vortex tangle regime when this number increases dramatically. The presence of anti-vortices is evidenced by the evolution to turbulence, which is not possible without a mixture of both types of vortices. In fact, if only one type of vortex were present, the cloud would evolve to a vortex lattice. Also, we observe formation of vortex clusters \cite{seman} similar to what is theoretically predicted \cite{motonen} for the coexistence of vortices and anti-vortices in the sample.  

\subsection{Unusual free expansion}

\begin{figure}
\begin{center}
\includegraphics[%
  width=0.6\linewidth,
  keepaspectratio]{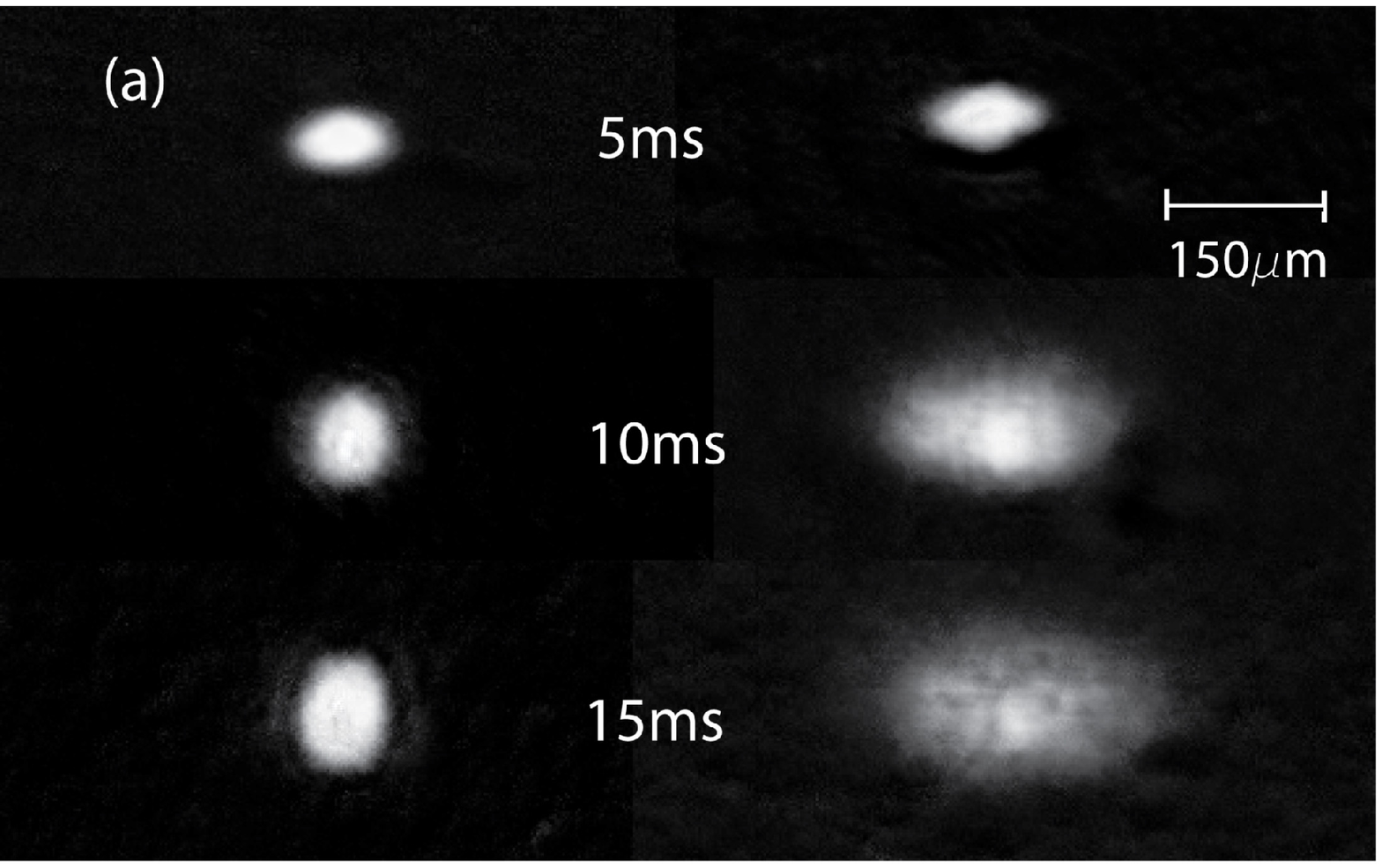}
 \includegraphics[%
  width=0.6\linewidth,
  keepaspectratio]{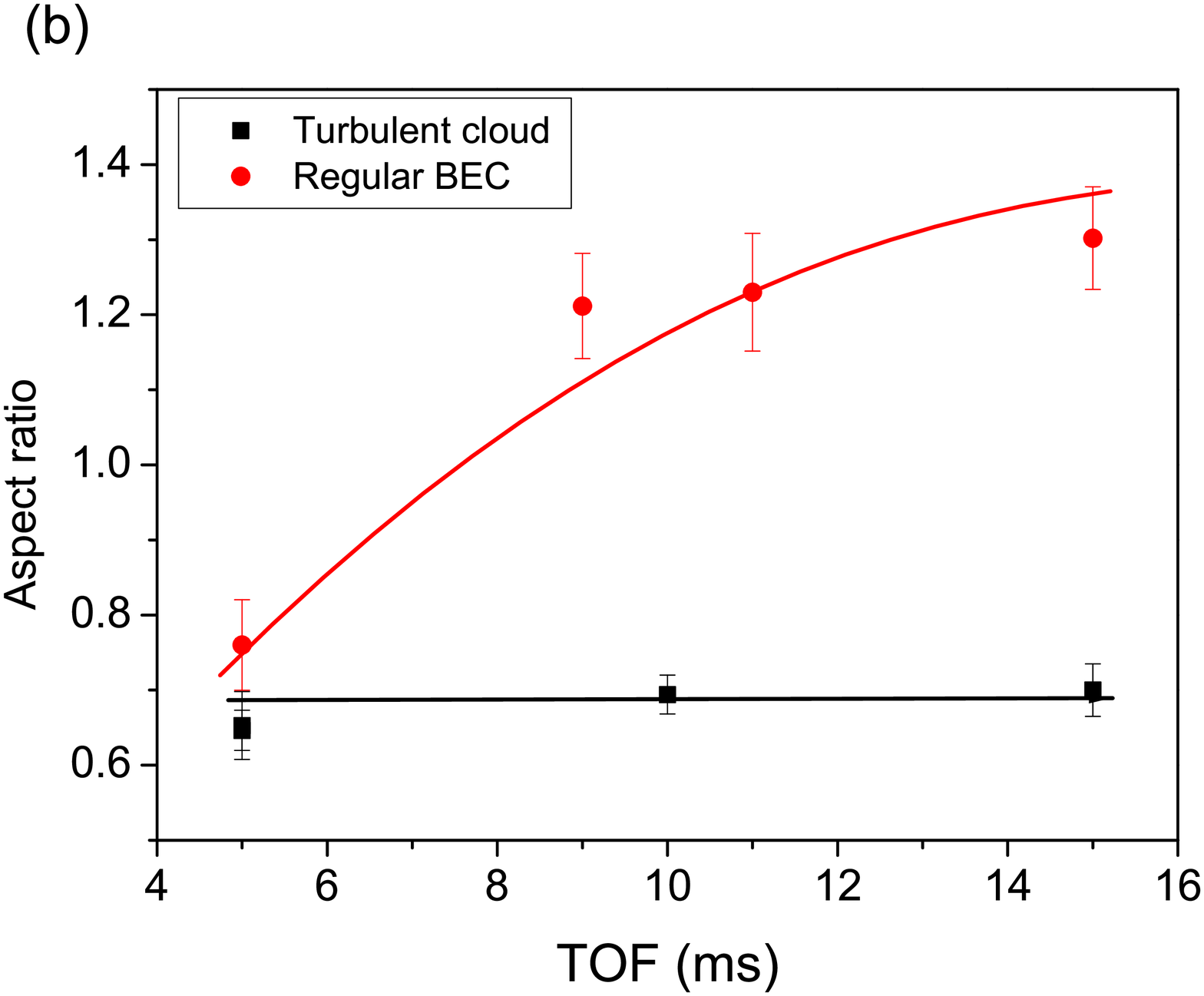} 
\end{center}
\caption{(Color online) (a) Side-by-side images of a regular, non-excited BEC and a turbulent cloud showing the aspect ratio inversion on the former case and the suppression of that inversion in the turbulent regime. (b) Aspect ratio (ration between main axis) of the BEC and turbulent clouds evidencing the inversion of the first and maintenance of the latter.}
\label{4}
\end{figure}

The appearance of tangled vortices is certainly a strong evidence of turbulence in our system. Nevertheless, as expected from any system that makes the transition from a regular flow to a turbulent flow, the hydrodynamic behavior of the cloud is also changed. That unusual behavior manifests itself as a suppression of the aspect ratio (ratio between the most and least confining axis of the cloud) inversion of the cloud. This can be seen clearly in Figs.\ref{4}a and \ref{4}b where we place side by side an expanding non-excited BEC where the aspect ratio inversion is seen and a turbulent cloud that expands keeping its aspect ratio constant. Fig.\ref{4}b is the numerical evaluation of the aspect ratio of the expanding clouds as a function of time. It is important to emphasize that a non-condensed cloud expands towards isotropy or, in other words, it starts with an anisotropic shape (the shape of the confining potential), but it expands and reaches unitary aspect ratio, does not going beyond. Those facts together let us conclude that the turbulent cloud keeps a quantum behavior while expanding, though this behavior is not the expected one for a quantum degenerate bosonic cloud.

In brief, the equation that governs the behavior of a BEC is the Gross-Pitaevskii equation, given by \begin{equation} \left[-\frac{\hbar^2\nabla^2}{2m}+V_{ext}(\vec{r})+U_0\left |\Psi(\vec{r},t)\right |^2\right]\Psi(\vec{r},t)=i\hbar\frac{\partial}{\partial t}\Psi(\vec{r},t)\label{eq:TDGPE},\end{equation} where $-\frac{\hbar^2\nabla^2}{2m}$ is the kinetic energy, $V_{ext}(\vec{r})$ is the confining potential and $U_0\left |\Psi(\vec{r},t)\right |^2$ accounts for the interaction energy.
In a expanding condensate, $V_{ext}(\vec{r})=0$ and the whole expansion is governed by the interplay of kinetic and interaction energies. For a thermal cloud the kinetic term dominates and since it is isotropic the expansion is also isotropic. For a BEC we have the other way round behavior: $-\frac{\hbar^2\nabla^2}{2m}<<U_0\left |\Psi(\vec{r},t)\right |^2$, which means that the anisotropic interaction energy (coming from the anisotropic spatial distribution $\left|\Psi(\vec{r},t)\right |^2$) governs the expansion, giving rise to the aspect ratio inversion phenomena, one of the strongest signatures of quantum degeneracy in a system of bosons. In Ref.\cite{castin} authors present a detailed description of this problem for a $T=0$ bosonic cloud. For $T>0$ we refer to Ref.\cite{caracanhas}, where a modified Hartree-Fock treatment takes into account the effects of the thermal cloud on the expanding condensate. In any case, the aspect ration inversion of the condensate is predicted.
Under these considerations, we cannot explain the observed behavior of the turbulent cloud. There is certainly a kind of self-trapping due to the vortex tangle configuration that prevents the cloud to leave its initial spatial distribution. It may well be that the presence of a randomic distributed field of rotations prevents the atoms to turning into an isotropic expansion. The idea is that the rotational field distribution creates an effective potential that expands together with the cloud. We should also observe that the time of flight for the turbulent cloud indicates that a much hotter fluid is obtained. That can be an indicative from decay mechanisms of the QT regime. Formation of Kelvin waves during vortices reconnections may be present and producing energy release in the form of phonons that increase the overall cloud energy. The increase in temperature of the cloud is a topic currently under investigation. 

\section{Conclusions}

In this work we have described a technique no generate vortices and anti-vortices in a BEC with evolution to Quantum Turbulence. Turbulence manifests itself by the appearance of a vortex tangle in the sample in contrast to the regular vortices regime. The turbulent state shows up also in a dramatic change in the hydrodynamic behavior of the cloud, where a suppression of the aspect ratio inversion is observed. Instead, the cloud expands keeping its original spatial profile and constant aspect ratio.

\begin{acknowledgements}
We acknowledge financial support from FAPESP (CEPID program), CNPq (INCT program) and CAPES.
\end{acknowledgements}
%


\begin{thebibliography}{99}

\bibitem{Hall} H. E. Hall and W. F. Vinen, Proc. Royal Soc. London, \textbf{238}, 204 (1956)
\bibitem{QT}W. F. Vinen and J. J. Niemela, J. Low Temp. Phys. \textbf{128}, 167 (2002)
\bibitem{QT2}C. F. Barenghi, R. J. Donnelly, and W. F. Vinen (Eds.), Quantized Vortex Dynamics and Superfluid Turbulence, Springer Verlag (2001).
\bibitem{tsubotaturb}M. Kobayashi and M. Tsubota, Phys. Rev. A \textbf{76}, 045603 (2007)
\bibitem{Hemethod} M. Blazkova \textit{et al.}, Phys. Rev. B \textbf{79} 054522 (2009) and references therein
\bibitem{hydrogen} G. P. Bewley \textit{et al.}, Nature \textbf{441}, 588 (2006)
\bibitem{vfm} W. P. Halperin and M. Tsubota (Eds.), Progress in Low Temperature Physics, vols.\textbf{16}, Elsevier, Amsterdam (2008)
\bibitem{bec} M. H. Anderson \textit{et al.}, Science \textbf{269}, 198 (1995)
\bibitem{vortex} M. R. Matthews \textit{et al.}, Phys. Rev. Lett. \textbf{83}, 2498 (1999); K. W. Madison \textit{et al.},  Phys. Rev. Lett. \textbf{84} 806 (2000)
\bibitem{assymetric potential} E. Hodby \textit{et. al}, Phys. Rev. Lett. \textbf{88} 010405 (2001)
\bibitem{madison} S. Inouye \textit{et. al} Phys. Rev. Lett. \textbf{87}, 080402 (2001); K. W. Madison \textit{et. al}  Phys. Rev. Lett. \textbf{84} 806 (2000); F. Chevy \textit{et. al}, Phys. Rev. Lett. \textbf{85} 2223(2000). 
\bibitem{abo} J. R. Abo-Shaeer \textit{et. al}, Science \textbf{292}, 476 (2001)
\bibitem{vortexsimul} M. Tsubota \textit{et al.}, Phys. Rev. A \textbf{65}, 023603 (2002); K. Kasamatsu \textit{et al.}, Phys. Rev. A \textbf{67}, 033610 (2003); A. A. Penckwitt \textit{et al.}, Phys. Rev. Lett. \textbf{89}, 260402 (2002); E. Lundh \textit{et al.}, Phys. Rev. A \textbf{67}, 063604 (2003); C. Lobo \textit{et al.}, Phys. Rev. Lett. \textbf{92}, 020403 (2004).
\bibitem{Scherer}  D. R. Scherer \textit{et. al} Phys. Rev. Lett. \textbf{98} 110402 (2007).
\bibitem{LG} M. F. Andersen \textit{et. al} Phys. Rev. Lett. \textbf{97} (2006) 170406; K. C. Wright \textit{et. al} Phys. Rev. A \textbf{77} 041601(R) (2008).
\bibitem{BJP} E. A. L. Henn \textit{et al.}, Braz. Journ. Phys. \textbf{38}, 279 (2008)
\bibitem{vortexform} E. A. L. Henn \textit{et al.}, Phys. Rev. A \textbf{79}, 043618 (2009)
\bibitem{PRL} E. A. L. Henn \textit{et al.}, Phys. Rev. Lett. \textbf{103}, 045301 (2009).
\bibitem{caracanhas} M. Caracanhas \textit{et al.}, J. Phys. B: At. Mol. Opt. Phys. \textit{42}, 145304 (2009)
\bibitem{Muniz} S. R. Muniz \textit{et al.}, Phys. Rev. A \textbf{73}, 041605R (2006); S. R. Muniz \textit{et al.}, Math. and Comp. in Simul., \textbf{74}, 397 (2007)
\bibitem{insitu} M.R. Andrews \textit{et al.}, Science \textbf{273}, 84 (1996)
\bibitem{KH} G. E. Volovik, JETP Letters \textbf{75}, 418 (2002)
\bibitem{KH theory} S. E. Korshunov, JETP Lett. \textbf{75}, 423 (2002).
\bibitem{KH superfluid} R. Blaauwgeers, \textit{et al.}, Phys. Rev. Lett. \textbf{89} 155301 (2002)
\bibitem{seman} arXiv:0907.1584, submitted for publication (2009)
\bibitem{motonen} M. Mottonen \textit{et al.}, Phys. Rev. A \textbf{71}, 033626 (2005)
\bibitem{castin} Y. Castin and R. Dum, Phys. Rev. Lett. \textbf{77}, 5315 (1996)
\end{thebibliography}
\end{document}